\documentclass[10pt,twocolumn,preprintnumbers,amsmath,amssymb,nofootinbib,superscriptaddress,prd]{revtex4-2}
\usepackage[dvipsnames]{xcolor}
\usepackage{comment}
\usepackage{bm}
\usepackage{graphicx}
\usepackage{tabularx}
\usepackage[colorlinks,allcolors=RoyalBlue]{hyperref}
\usepackage[normalem]{ulem}
\usepackage{xspace}
\usepackage{hyperref,url}
\usepackage{orcidlink}

\newcommand{\chisq}{\chi^2}
\newcommand{\dchisq}{\ensuremath{\Delta\chi^2}}
\newcommand{\LCDM}{$\Lambda$CDM\xspace}
\newcommand{\lcdm}{$\Lambda$CDM\xspace}

\newcommand{\fede}{f_{\rm EDE}}

\newcommand{\hunitstwo}{\ensuremath{\mathrm{km}\,\mathrm{s}^{-1} \,\mathrm{Mpc}^{-1}}}
\newcommand{\planck}{\textit{Planck}\xspace}
\usepackage[capitalise]{cleveref}

\newcommand{\rev}[1]{{#1}}

\begin{document}
\title{Disentangling cosmic distance tensions with early and late dark energy}
\author{Tanisha Jhaveri}
\affiliation{Department of Astronomy and Astrophysics, University of Chicago, 5640 South Ellis Avenue, Chicago, IL, 60637, USA}
\affiliation{Kavli Institute for Cosmological Physics, Enrico Fermi Institute, University of Chicago, 5640 South Ellis Avenue, Chicago, IL, 60637, USA}
\author{Tanvi Karwal}
\affiliation{Department of Astronomy and Astrophysics, University of Chicago, 5640 South Ellis Avenue, Chicago, IL, 60637, USA}
\affiliation{Kavli Institute for Cosmological Physics, Enrico Fermi Institute, University of Chicago, 5640 South Ellis Avenue, Chicago, IL, 60637, USA}
\author{Thomas Crawford\,\orcidlink{0000-0001-9000-5013}}
\affiliation{Department of Astronomy and Astrophysics, University of Chicago, 5640 South Ellis Avenue, Chicago, IL, 60637, USA}
\affiliation{Kavli Institute for Cosmological Physics, Enrico Fermi Institute, University of Chicago, 5640 South Ellis Avenue, Chicago, IL, 60637, USA}
\author{Wayne Hu}
\affiliation{Department of Astronomy and Astrophysics, University of Chicago, 5640 South Ellis Avenue, Chicago, IL, 60637, USA}
\affiliation{Kavli Institute for Cosmological Physics, Enrico Fermi Institute, University of Chicago, 5640 South Ellis Avenue, Chicago, IL, 60637, USA}
\author{Ali Rida Khalife\,\orcidlink{0000-0002-8388-4950}}
\affiliation{Sorbonne Universit\'e, CNRS, UMR 7095, Institut d'Astrophysique de Paris, 98 bis bd Arago, 75014 Paris, France}
\author{Lennart Balkenhol\,\orcidlink{0000-0001-6899-1873}}
\affiliation{Sorbonne Universit\'e, CNRS, UMR 7095, Institut d'Astrophysique de Paris, 98 bis bd Arago, 75014 Paris, France}
\author{Fei Ge}
\affiliation{California Institute of Technology, 1200 East California Boulevard., Pasadena, CA, 91125, USA}
\affiliation{Kavli Institute for Particle Astrophysics and Cosmology, Stanford University, 452 Lomita Mall, Stanford, CA, 94305, USA}
\affiliation{Department of Physics, Stanford University, 382 Via Pueblo Mall, Stanford, CA, 94305, USA}
\affiliation{Department of Physics \& Astronomy, University of California, One Shields Avenue, Davis, CA 95616, USA}

\begin{abstract}
Recent cosmological data reveal tension between parameters inferred from measurements of the cosmic microwave background (CMB), baryon acoustic oscillations (BAO), and supernovae (SN) under $\Lambda$CDM. Typical dynamical dark energy parameterizations (such as $w_0w_a$) that seek to jointly resolve these tensions have an equation of state parameter that crosses into the phantom regime, leading to potential instabilities for physical models.  We show that the BAO (early-time) and SN (late-time) sides of the tension can instead be treated independently. Early dark energy (EDE) can reduce the tension between CMB-BAO data by changing the calibration of the sound horizon at the drag epoch $r_d$, with a $\Delta\chi^2 = -\rev{9.4}$ relative to $\Lambda$CDM, raising $H_0$ to \rev{70.87} \hunitstwo.  EDE alone cannot bring consistency between CMB, BAO, and SN data, but combining with a thawing-quintessence component of dark energy reduces \rev{tensions} between the three datasets, with \rev{$\Delta\chi^2=-12.6$ relative to $\Lambda$CDM without 
a phantom component, vs.\ $\Delta\chi^2=-15.8$ for $w_0 w_a$} with one.
We consider \rev{different SN datasets, using the most recent DES Dovekie catalog as our default while assessing differences with the original DESY5 and Pantheon+ catalogs}. While the significance of adding thawing quintessence changes, the EDE solution to the CMB-BAO tension remains nearly unaffected. 
\rev{Moreover, though we do not include direct Hubble constant measurements in these $\Delta\chi^2$ values, the EDE solution reduces the Hubble tension with the Local Distance Network value from $7\sigma$ in $\Lambda$CDM to $2-3\sigma$ depending on the SN dataset, nominally the equivalent of an extra $\Delta\chi^2 \sim -40$ or more.}
\end{abstract}

\date{\today}

\maketitle

\section{Introduction} 
\label{sec:intro}

The standard \LCDM model of cosmology has proven enormously successful in fitting numerous cosmological data sets across a wide range of physical scales, redshifts, observing wavelengths, and cosmological probes (see~\cite{Turner:2022gvw} for a recent review). Perhaps the most notable success of \lcdm\ is in fitting the complex form of the angular power spectra of temperature and polarization anisotropies in the cosmic microwave background (CMB). Recent results from precision CMB experiments continue to find consistency with this simple six-parameter model of cosmology. For example, the most recent compilation of CMB data from the \planck\ satellite \cite{Planck:2018nkj} and the ground-based Atacama Cosmology Telescope (ACT, \cite{Henderson:2015nzj}) and South Pole Telescope (SPT, \cite{Carlstrom:2009um}) demonstrates that the three data sets are mutually consistent and consistent with \lcdm,
despite being tested by data at very different angular scales whether in temperature or polarization, and despite the enormous constraining power of each of the data sets \cite{SPT-3G:2025bzu}.

However when CMB constraints are compared with those from other observables, small but statistically significant discrepancies have begun to arise in recent years, the most significant of these being the Hubble tension. 
 Combining the most precise CMB probes results in a best-fit value under \lcdm of the current expansion rate $H_0 = 67.24 \pm 0.35$ \cite{SPT-3G:2025bzu}, while the Local Distance Network (H0DN) collaboration
 reports a direct measurement of $H_0 = 73.50 \pm 0.81$ \cite{H0DN:2025lyy}, largely but not exclusively based on the SH0ES Cepheid-supernova Ia local distance ladder \cite{Riess:2021jrx,Riess:2025chq}.  Here and throughout, the units of $H_0$ values are implicitly  $\hunitstwo$. Taken at face value, this is a 
 7.1$\sigma$ discrepancy, with a vanishingly small associated probability of arising from statistical fluctuations alone. The CCHP group, using a different calibration of the distance ladder and a different SN sample, 
finds lower statistical significance for this tension, $H_0= 70.39\pm 1.9$
\cite{Freedman:2024eph}, where we have added their statistical and systematic errors in quadrature. 
It is currently unclear whether this discrepancy indicates new physics or unaccounted-for systematics in the analysis of any of the data sets.

More recently, other tensions have begun to grow between the $\Lambda$CDM parameters preferred by three  key cosmological probes: the CMB, the baryon acoustic oscillation (BAO) signature in spectroscopic galaxy surveys, and the distance-redshift relation inferred from supernova (SN) Ia data. This mutual tension has been interpreted most widely as a preference for dynamical dark energy, which reaches a  significance of $4.2 \sigma$ \cite{DESI:2025zgx} using a combination of CMB data from \planck, BAO measurements from the second data release of the Dark Energy Spectroscopic Instrument (DESI-DR2, \cite{DESI:2025zgx}), and SN distances from the 5-year Dark Energy Survey (DES-Y5) data set \cite{DES:2024jxu}, though the significance drops \rev{to $3.2\sigma$} with the most recent DES Dovekie revision \cite{DES:2025sig}.
From a model-building standpoint, it is notable that the Chevallier-Polarski-Linder \cite{Chevallier:2000qy, Linder:2002et} or $w_0w_a$ parameterization for evolving dark energy used in Ref.\ \cite{DESI:2025zgx} results in a best-fit dark energy equation of state that evolves from the normal ($w\ge -1$) to the phantom ($w<-1$) regime between the redshifts best measured by SN, BAO and the CMB. This phenomenology 
places strong requirements on physical models in order not to exhibit ghost and gradient instabilities, which typically requires fundamental modifications to gravitational or dark sector forces
\cite{Vikman:2004dc,Hu:2004kh,Creminelli:2008wc,Fang:2008sn,Gubitosi:2012hu,Bloomfield:2012ff,Amendola:1999er,Das:2005yj,Copeland:2006wr} (see also 
\cite{Oliveira:2025uye,Pullisseri:2025ran,Wolf:2024stt,Wolf:2025jed,Wolf:2025acj,Pourtsidou:2025sdd,Tsujikawa:2025wca,Yao:2025wlx,Guedezounme:2025wav,Hogas:2025ahb,Paliathanasis:2025hjw,Andriot:2025los,You:2025uon,Pan:2025qwy,Silva:2025hxw,SanchezLopez:2025uzw,Baryakhtar:2024rky,Bottaro:2024pcb,Costa:2025kwt,Khoury:2025txd,Bedroya:2025fwh,Caldwell:2025inn,Weiner:2026sfm} for recent assessments).
 
Furthermore, the dynamical dark energy solution---and most proposed solutions---to the CMB-BAO-SN tension does nothing to alleviate the $H_0$ tension, and indeed makes it worse if phantom crossing is forbidden.
The tension between CMB and BAO mainly involves the calibration of their standard rulers---the sound horizon at recombination and the end of the Compton drag epoch, respectively---and the corresponding implications for the relative distance between BAO redshifts and recombination.
This tension can be relaxed in various ways that do not involve dark energy, such as raising the optical depth through reionization
\cite{Sailer:2025lxj,Jhaveri:2025neg} (cf.~\cite{Cain:2025usc,Garcia-Gallego:2025jrn} for limits from other reionization observables)
or adding a small spatial curvature to the background \cite{DESI:2025zgx,Chen:2025mlf}. \rev{Meanwhile,} the larger low- vs.\ high-$z$ SN distance moduli compared to the predictions of the best-fit CMB+BAO $\Lambda$CDM model
can be simultaneously resolved by a quintessence component of non-phantom dark energy
\cite{Liu:2025bss}.  These solutions, while successfully avoiding the phantom dark energy of the $w_0 w_a$ model,  fail to jointly address the Hubble tension.

By contrast, the early dark energy (EDE) model, an extension to \lcdm\ originally proposed to resolve the Hubble tension
\cite{Karwal:2016vyq,Poulin:2018cxd,Poulin:2023lkg,Kamionkowski:2022pkx, Smith:2019ihp} 
(see also 
\cite{Agrawal:2019lmo,Lin:2019qug,Alexander:2019rsc,Sakstein:2019fmf,Das:2020wfe,Niedermann:2019olb,Niedermann:2020dwg,Niedermann:2021vgd,Ye:2020btb,Berghaus:2019cls,Freese:2021rjq,Braglia:2020bym,Sabla:2021nfy,Sabla:2022xzj,Gomez-Valent:2021cbe,Moss:2021obd,Guendelman:2022cop,Karwal:2021vpk,McDonough:2021pdg,Wang:2022nap,Alexander:2022own,McDonough:2022pku,Nakagawa:2022knn,Gomez-Valent:2022bku,MohseniSadjadi:2022pfz,Kojima:2022fgo,Rudelius:2022gyu,Oikonomou:2020qah,Tian:2021omz,Maziashvili:2021mbm,Berghaus:2022cwf,Ramadan:2023ivw,Bansal:2025usn, Garny:2025kqj, Guendelman:2025swp, Smith:2025grk, Bisabr:2025qkt, Kamionkowski:2024axz, Simon:2024jmu, Sohail:2024oki, Garny:2024ums}
for variants on this original model), 
can also ease the CMB-BAO tension \cite{Poulin:2025nfb,ACT:2025tim,SPT-3G:2025vyw}.\footnote{The  \texttt{P-ACT-LB} result from \cite{ACT:2025tim}, using BAO data from DESI DR1, showed  that the reduction in tension \rev{ was only $\Delta \chi^2_{\rm EDE} - \Delta \chi^2_{\Lambda \rm CDM} = -6.6$} for that data set, and the authors of \cite{ACT:2025tim} described the result as not statistically significant though the significance should increase if DESI DR1 data were replaced with DESI DR2 as we do here.}
EDE is usually realized as a scalar field that begins to roll in the early universe, close to matter-radiation equality, and quickly dilutes away, such that its impact on cosmology is localized in redshift to a short-lived boost in the expansion rate before recombination. 
This shrinks the size of the sound horizon and changes the relative distance from $z\sim 1$ to recombination, easing the tension with DESI BAO and simultaneously increasing the CMB-inferred value of $H_0$, diminishing the Hubble tension (see also \cite{Lynch:2024hzh,SPT-3G:2025bzu} for alternatives involving modified recombination).

The potential role of EDE in alleviating both the Hubble tension and the three-way tension between CMB, BAO, and the Pantheon+ SN data \cite{Brout:2022vxf} was studied recently using \planck\ and ACT \cite{Poulin:2025nfb}, and including the latest SPT data \cite{SPT-3G:2025vyw,Wang:2025djw}.
Since EDE does not directly change the SN distance moduli at late times, the more SN data favor a quintessence component, the worse the overall fit to CMB, BAO and SN will be with EDE alone.  
We expand on these studies, incorporating profile-likelihood constraints, investigating the impact of 
using different SN data sets, in particular \rev{the most recent DES 
Dovekie release \cite{DES:2025sig} as well as the original DESY5 release \cite{DES:2024jxu} 
which most strongly favors quintessence over $\Lambda$CDM}, and explicitly testing the viability of non-phantom thawing quintessence models in conjunction with EDE (see also \cite{Reboucas:2023rjm} for a pre-DESI exploration).

Profile-likelihood studies are especially important for 
EDE models because they are susceptible to prior-volume effects \cite{Smith:2020rxx,Herold:2021ksg,Poulin:2023lkg,Karwal:2024qpt}: as the amount of EDE becomes negligible, its other parameters become unconstrained by the data, leading to a large prior volume in the \lcdm limit. 
This ballooning prior volume can pull Bayesian posteriors towards this region and away from the peak in the likelihood. 
Profile likelihoods are a prior-independent statistical approach that trace the data likelihood along a parameter direction, providing complementary constraints to the Bayesian posteriors. We examine both Bayesian posteriors and profile likelihoods for the main data and model combinations in this work and discuss the role of prior-volume effects in interpreting the results.

\rev{
This paper is organized as follows. We discuss the methodology in Sec.~\ref{sec:methodology} focusing on dark energy models, datasets and statistical methods.  We address CMB-BAO tension resolution in the EDE model and discuss its implications for the Hubble tension in Sec.~\ref{sec:cmbbao}.  We consider the impact of adding the Dovekie SN data  and alternates on these solutions with and without a thawing quintessence component in Sec.~\ref{sec:sn} and discuss these results in Sec.~\ref{sec:discussion}.
}

\section{Methodology}
\label{sec:methodology}

In this section, we summarize the models, data, and statistical methods used in our analysis.

\subsection{Early and Late Dark Energy Models}

Early dark energy (EDE) models introduce an additional component to the Universe that impacts the expansion rate around matter-radiation equality.   
The canonical EDE model is an axion-like scalar field that is initially frozen at some value $\phi_i=\theta_i f_\phi$ on its potential 
\begin{equation}
    V(\phi) = m_\phi^2f_\phi^2[1-\cos(\phi/f_\phi)]^3 \,,
\end{equation}
until the Hubble rate drops sufficiently so that it can oscillate and decay to its minimum.  
Here $m_\phi$ and $f_\phi$ are the mass and decay constant of the axion \rev{respectively}.
The scalar field energy density $\rho_\phi$ thus initially behaves like a cosmological constant $\rho_\phi= $\,const.\ and later dilutes faster than matter or radiation as $\rho_\phi \propto a^{-9/2}$, where $a$ is the scale factor. 
Relative to other $\Lambda$CDM components, EDE is therefore only important at a single characteristic epoch. 
The theory parameterization of EDE based on $\{\phi_i, m_\phi, f_\phi\}$ is generally translated into a phenomenological parameterization of $\{\fede, z_c, \theta_i\}$, where $\fede$ is the peak fractional energy density of EDE occurring at its critical redshift $z_c$, given $\theta_i$, its initial field position in units of $f_\phi$.

For late dark energy, we employ the phenomenological but flexible Chevallier-Polarski-Linder \cite{Chevallier:2000qy, Linder:2002et} parametrization of the time-varying dark energy equation of state parameter:
\begin{equation}
    w(a) = w_0 + w_a(1-a) \,,
\label{eq:w0wa}
\end{equation}
which 
can reproduce
the observational impact during the acceleration epoch of a wide range of physical models.
When $w_0$ and $w_a$ are allowed to take on any value,
the CMB-BAO-SN tension favors a combination where $w(a)$
evolves across $w=-1$.  
To enforce stability below the horizon for such cases, which is difficult for physical models, we instead use the Parameterized Post-Friedmann approach for perturbations 
\cite{Fang:2008sn}.
In this context, we refer to such solutions as $w_0 w_a$ models.   

Scalar field quintessence models of dark energy are manifestly stable but since $w\geq-1$, \rev{they} cannot cross the phantom divide. 
The simplest and best motivated models, where the scalar field is frozen on its potential by Hubble drag and then released, similarly to EDE, are called thawing quintessence models.  Their observational impact can be mimicked by \cref{eq:w0wa} with \rev{the fitted form,}
\begin{equation}
    w_a = -1.58(1+w_0) \,,
\end{equation}
leaving $w_0$ as the single free parameter controlling the dynamics \rev{(see Eq.\ (14) of \cite{DESI:2025fii})}.\footnote{Although this form still allows $w<-1$ as $a\rightarrow 0$, this form nonetheless matches distance measures for thawing quintessence models \cite{dePutter:2008wt,Linder:2024rdj}.}
We call these models calibrated thawing quintessence models or ``thaw" for short.

We implement these models in a modified version of the \verb|CLASS| code, \verb|CLASS_EDE|,\footnote{\url{https://github.com/mwt5345/class_ede}}  which incorporates the canonical EDE component and a dark energy fluid. 
Our baseline model varies the six \lcdm parameters: the physical densities $\omega_b$ of baryons and $\omega_c$ of cold dark matter, 
the angular size $\rev{\theta_s}$ of the sound horizon,\footnote{\rev{CLASS-based codes use $\theta_s$ for the angular size of the sound horizon $r_s$ as defined by the peak of the visibility function, unlike the \planck convention where $\theta_*$ is the angular size of the sound horizon ${r_*}$ defined by  optical depth unity during recombination, i.e.\ excluding reionization.   The CLASS derived parameter, also called $\theta_*$ is defined by total optical depth unity including reionization and should not be confused with the more standard definition.}}
the amplitude $\ln A_s$ and tilt $n_s$ of the primordial power spectrum and the optical depth $\tau$ to reionization. 
These parameters have their usual flat, uninformative priors. In addition, we vary the EDE parameters 
with flat priors in the ranges
\begin{equation}
    \fede \in [0,0.3], \quad
    \log_{10}z_c \in [3,4.3], \quad
    \theta_i \in [0.1,3.1] \,.
\end{equation}
When considering late dark energy, we vary 
$\{w_0 \,, w_a\}$ or $\{ w_0 \}$, depending on the extension, with priors 
\begin{equation}
    w_0 \in[-3,1], \quad
    w_a \in[-3,2] \,.
\end{equation}
Neutrinos are fixed at minimal mass for normal ordering $\Sigma m_\nu = 0.06$ eV, with \rev{a single massive species}.\footnote{\rev{Due to CMB+BAO tension, alternate choices that are indistinguishable with CMB, such as 3 degenerate-mass neutrinos, can make a $\Delta\chi^2\lesssim 1$ difference in CMB+BAO; $\Lambda$CDM, which carries through to model comparisons.}}
From these fundamental parameters, \rev{we derive others that reveal tensions across datasets.
These include $H_0$, $\Omega_m$ (total matter density today in units of the critical density), and the ``weak-lensing amplitude parameter'' $S_8 \equiv \sigma_8 (\Omega_m/0.3)^{0.5}$. Here $\sigma_8$ is the current value of the RMS linear density fluctuation in spheres of radius $8 \ \mathrm{Mpc}/h$. }

Certain precision parameters are recommended by Ref. \cite{ACT:2025tim} (their Fig. 48) which increase the precision of Boltzmann codes for comparing with high-resolution CMB surveys like ACT and SPT. 
We adopt a subset of these parameters, balancing speed and accuracy, choosing \rev{higher accuracy only when the precision setting has a $\dchisq > 0.2$ impact on the individual likelihoods considered here}, checked at multiple points in \lcdm and EDE cosmologies. 
This subset of parameters is shown in \cref{tab:precisions}. 

\begin{table}
   \centering
   \begin{tabular}{|c|c|}
   \hline
       \textbf{\texttt{CLASS} argument} & \textbf{Value} \\
       \hline
       \verb|non_linear|   & \verb|hmcode| \\
       \verb|l_linstep|   & $20$ \\
       \verb|l_max_ur|   & $35$ \\
       \verb|transfer_neglect_delta_k_S_t0|        & $0.17$ \\
       \verb|transfer_neglect_delta_k_S_e|        & $0.17$ \\
   \hline
   \end{tabular}
   \caption{Settings for \texttt{CLASS} precision parameters beyond the defaults. }
   \label{tab:precisions}
\end{table}

\subsection{Datasets}
\label{sec:data}

\rev{We select for our default analysis the following datasets that contribute tension in $\Lambda$CDM:}

\begin{itemize}
    \item CMB primary anisotropy: We use the
        latest data from the Atacama Cosmology Telescope (ACT DR6, \cite{ACT:2025fju}) and the South Pole Telescope (SPT-3G D1, \cite{SPT-3G:2025bzu}), and the \planck\ 2018 data. For \planck, we use the PR3 low-$\ell$ TT and EE likelihoods and the \texttt{Plik} high-$\ell$ TTTEEE likelihoods \cite{Planck:2018vyg}. Following \cite{ACT:2025fju}, we reduce the covariance between ACT DR6 and \planck\ (both of which cover a significant fraction of sky) by using only \planck\ data at $\ell<1000$ in TT and $\ell<600$ in TE and EE, and using only ACT DR6 data above those multipole values. Because of the small sky overlap between SPT-3G D1 and ACT/\planck, we use the full multipole range of the SPT-3G D1  likelihood, as in \cite{SPT-3G:2025bzu}. \rev{We use \texttt{lite} likelihoods as supplied by the \planck\ collaboration and by \texttt{candl} \cite{Balkenhol:2024sbv}}.\footnote{\rev{Lite likelihoods \cite[e.g.,][]{Dunkley:2013vu} are marginalized over foregrounds using the whole dataset of each experiment and so are both robust and conservative when combining cut versions of each. We find in particular that using multifrequency \planck\ data instead with our cuts can lead to inappropriately unconstrained point source amplitudes in particular, allowing for more freedom in the cosmology models and can affect $\Delta\chi^2$ between models by ${\cal O}(1)$.}}
    \\
    \\
    CMB lensing: We use the combined \planck\ PR4 lensing \cite{Carron:2022eyg} and ACT DR6 lensing \cite{ACT:2023dou} datasets, accounting for their cross-correlations. To these, we add SPT-3G MUSE lensing data \cite{SPT-3G:2024atg} as an independent dataset and neglect cross-correlations which are negligible due to the small but deep sky coverage \cite{ACT:2025rvn}. 
    \item BAO: We use the DESI DR2 likelihood 
    based on Table \rev{IV} in  Ref.~\cite{DESI:2025zgx} for $D_V/r_d$ for the lowest redshift bin and $D_M/r_d$, $D_H/r_d$, and their cross-correlation $r_{M,H}$ for all other bins.\footnote{
    The DESI DR2 BAO and DES Y5 and Dovekie SN likelihoods employed here are available at \url{https://github.com/tkarwal/cosmo_likelihoods}}
    \rev{Here $D_{X}$ are  BAO-measured distances  and $r_d$ is the sound horizon at the end of the drag epoch (see \cref{sec:cmbbao}). }
  
    \item SN: We use distance moduli from DES \rev{Dovekie \cite{DES:2025sig}}
    as our baseline unless explicitly mentioned otherwise.
    The absolute magnitude $M$ is marginalized in posteriors and maximized in profiles.  The likelihood employs individual SN but for visualization purposes we \rev{use} the binned data after correcting the apparent magnitudes for peculiar velocities.

\end{itemize}

\rev{While we choose the most recent DES Dovekie catalog as our default SN dataset, we also} explore the impact on our most important results of replacing this choice with alternative SN datasets:
\begin{itemize}

\item SN ($z>0.1$): 
    Dovekie catalog but excluding non-DES SN, effectively removing SNe with $z<0.1$. 

\item SN (DESY5):  \rev{The original DESY5 catalog where the preference for $w_0w_a$ is the strongest
\cite{DES:2024jxu}.}

\item SN (Pantheon+): We take the distance moduli from \cite{Brout:2022vxf} with a likelihood where $M$ is explicitly marginalized over.

\end{itemize}
\rev{We choose not to include Union3, which has not been revised for the improved Dovekie calibration technique.  The existing Union3 catalog would  lie in between the DESY5 and Pantheon+ cases in its dynamical dark energy preference \cite{DESI:2025zgx}.}

\subsection{Statistical Methods}

Our approach to constraints is two-fold. 
We obtain Bayesian posteriors via a Metropolis-Hastings Markov chain Monte Carlo (MCMC) analysis, as is standard in cosmology, using \verb|MontePython|
\cite{Audren:2012wb,Brinckmann:2018cvx}\footnote{\rev{We have updated the MontePython v3.6.1  big-bang-nucleosynthesis consistency tables from the default 2007 values to the most recent 2025 values.}}
to explore the parameter space and \verb|GetDist| \cite{Lewis:2019xzd} to analyze the posteriors. 
We run chains until the Gelman-Rubin convergence statistic \cite{1992StaSc...7..457G}  reaches \rev{$R-1 \lesssim 0.03$}. 
While this is weaker than typical for a $\Lambda$CDM analysis, this is still a stringent criterion for EDE given the prior-volume effects that slow MCMC convergence. 
Our aim here is to determine the subset of data combinations and \lcdm extensions under which EDE is preferred in Bayesian posteriors despite its prior-volume effects rather than a full exploration of tails of the posterior. 

As a complement, we also obtain the $\chi^2$ at maximum likelihood  using \verb|Procoli| \cite{Karwal:2024qpt} for every data combination and model we explore.\footnote{\rev{Data-driven priors on nuisance parameters are included by \texttt{MontePython} as part of the reported likelihood and $\chi^2$.  These are $A_{\rm Pl}=1\pm 0.003$ for the calibration of \planck and ACT ($A_{\rm ACT}=A_{\rm Pl}$) and separately $T_{\rm cal}=1\pm 0.0036$ for SPT as recommended by Ref.~\cite{SPT-3G:2025bzu}.
}}
Using this, we report \dchisq, or relative goodness of fit compared to a \lcdm model, 
\begin{equation}
    \Delta \chi^2 = (\chi^2_{\rm model} - \chi^2_{\Lambda \rm CDM}) \Big|_{\rm dataset} \,,
\end{equation}
and these values constitute our main results.
With this frequentist approach, our aim is to address prior-volume effects impacting the Bayesian posteriors by finding the true best-fit points and their corresponding improvements in fitting data. 
This helps pinpoint whether tensions between data sets in \lcdm are resolved by any particular extensions, and how the changes in $\chisq$ are distributed across data sets.

We use the sign convention that negative \dchisq\ values represent improvements.  When highlighting whether a given improvement in \dchisq\ is significant given extra parameters $\Delta N$ in the fit, we sometimes quote the change in the Akaike Information Criteria \cite{Akaike:1974vps} of $\Delta{\rm AIC}=\dchisq-2\Delta N$.
For each data combination, we report \dchisq\ 
for every model relative to a \lcdm fit to that specific data combination.
As these \lcdm cosmologies are different for each combination, 
we also report the relative \dchisq\ of the best-fit \lcdm cosmology to the CMB datasets alone as a reference, which can be used to evaluate the impact of any compromises in goodness of fit to the CMB when adding other datasets.
We refer to this model below as ``fiducial $\Lambda$CDM," and the parameter values for this model are shown in \cref{tab:pl18}. 
We reserve ``$\Lambda$CDM" to refer to the joint fit to a combination of data that may be in tension with the CMB. Furthermore, when \rev{comparing measured values of observables} against models,
we adopt this fiducial cosmology as the baseline.   
When observables are quoted in units of their values in the fiducial model, we use the notation [fid].  
For example,  since the CMB+BAO; EDE model in Tab.\ \ref{tab:pl18} gives \rev{$r_d=142.51$\,Mpc }
compared with 
\rev{$r_d=147.00 $\,Mpc}
for the fiducial $\Lambda$CDM model, we denote the EDE value as
\rev{$r_d$ [fid] $= 0.9694$}.

\begin{table}[]
    \centering
    \begin{tabular}{|c|c|c|}
    \hline
        \textbf{Parameter} & \textbf{Fid. $\Lambda$CDM} & \textbf{CMB+BAO; EDE} \\
    \hline
        $\omega_b$ & \rev{0.022390} & \rev{0.022606} \\
        $\omega_c$ & \rev{0.12034} & \rev{0.12736} \\
        $100\rev{\theta_s}$ & \rev{1.041605} & \rev{1.04125} \\
        $\rm ln(10^{10}A_s)$ & \rev{3.0488} & \rev{3.0716} \\
        $n_s$ & \rev{0.96766} & \rev{0.98393} \\
        $\tau$ & \rev{0.0555} & \rev{0.0611} \\
    \hline
        $f_{\rm EDE}$ & - & \rev{0.09} \\
        $\log_{10} z_c$ & - & \rev{3.54} \\
        $\theta_i$ & - & \rev{2.76} \\
    \hline
        $\chi^2_{\rm CMB}$ & \rev{\textbf{1000.8}} &   \rev{\textbf{999.3}} \\
        $\chi^2_{\rm BAO}$ & \rev{36.6} & \rev{\textbf{11.9}} \\
        $\chi^2_{\rm SN}$ & \rev{1640.9} & \rev{1644.4} \\
    \hline
        $\chi^2_{{\rm SN} (z > 0.1)}$ & \rev{1425.8} & \rev{1425.5} \\
        $\chi^2_{\rm SN (\rev{DESY5})}$ & \rev{1644.3} & \rev{1650.3} \\
        $\chi^2_{\rm SN (Pantheon+)}$ & \rev{1410.5} & \rev{1413.3} \\
    \hline
    \end{tabular}

    \caption{
    Parameter values and $\chi^2$ contributions of datasets under various models at the maximum likelihood point.  
    Fiducial $\Lambda$CDM only optimizes the CMB $\chi^2_{\rm CMB}$ whereas CMB+BAO; EDE optimizes the sum of $\chi^2_{\rm CMB}$ and $\chi^2_{\rm BAO}$, with the optimized contributions shown in boldface in each case. 
    We report results from data sets that are not included in the optimization to provide an absolute measure of tensions and their resolution.
    }
    \label{tab:pl18}
\end{table}

Finally, we further use \verb|Procoli| to evaluate the full profile likelihoods for a subset of our parameters \rev{for the relevant} model and dataset choices. In particular, we plot $\dchisq$ profiles for $\fede$ and $H_0$ in the EDE model with CMB+BAO data and the EDE+thaw model with CMB+BAO+SN data.
With $\fede$ profiles, we quantify the ability of EDE to address the CMB-BAO tension and the impact that SN have on this resolution.  With $H_0$ profiles, we quantify to what extent the Hubble tension is simultaneously resolved by EDE.

\section{CMB-BAO Tension and EDE}
\label{sec:cmbbao}

\begin{figure*}[ht] 
    \centering
    \includegraphics[width=\textwidth]{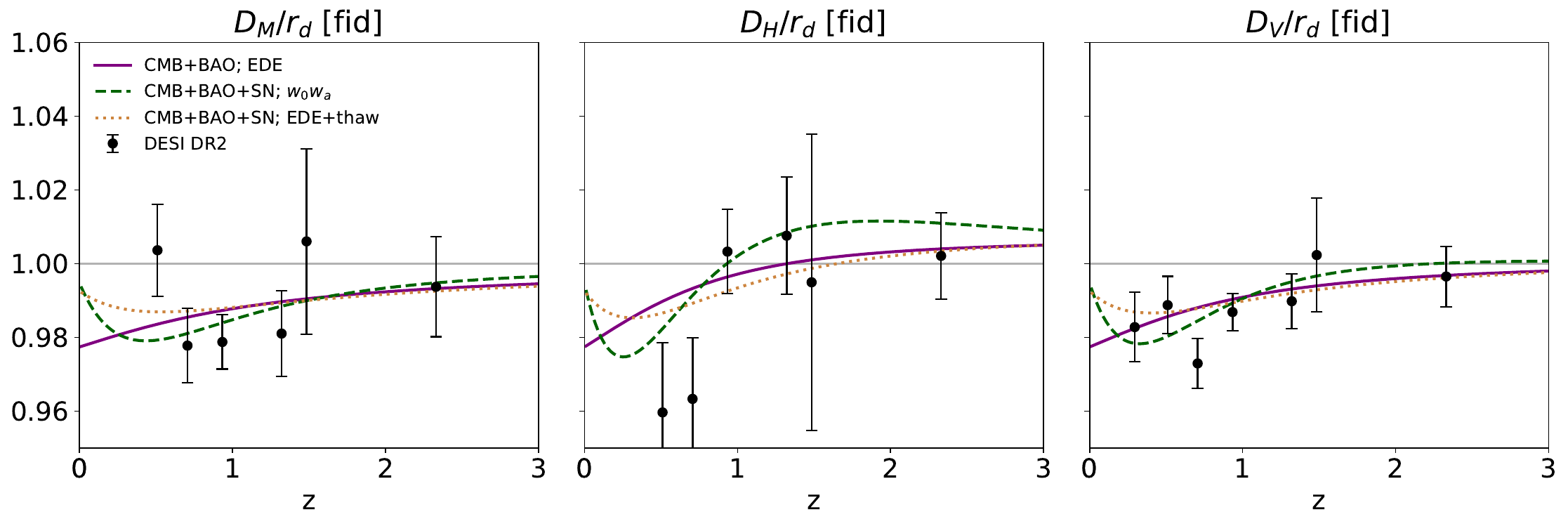}
    \caption{DESI DR2 BAO data relative to fiducial \lcdm $(D_{M,H,V}/r_d [{\rm fid}]\equiv 1)$. 
    BAO data prefer $1-2\%$ lower $D_M/r_d$ at $z\gtrsim 0.5$. 
    The curves show best-fit predictions of various models fit to different dataset combinations that relax the CMB-BAO tension. 
    With CMB+BAO; EDE (solid purple curve), the tension is eased by lowering $r_d$\rev{, which results in a higher value of} $H_0$.  
    This solution, which lowers $D_M/r_d$ at $z\lesssim 0.1$, is disfavored by SN (cf.~\cref{fig:SN}) with EDE alone, but the fit can be improved by pairing EDE with a thawing quintessence model of late dark energy (EDE+thaw, dotted orange curve). 
    This reconciliation among CMB+BAO and low-$z$ SN can also be achieved by $w_0w_a$ models (dashed green curve) but only through a phantom dark energy component.
    }
    \label{fig:bao_data}
\end{figure*}

We first address the tension between CMB and BAO data sets, which within $\Lambda$CDM is $>2\sigma$ \cite{DESI:2025zgx, SPT-3G:2025bzu, DESI:2025gwf}, the ability of EDE models to resolve these tensions, and the added benefit in these models of alleviating the Hubble tension.
Tension within $\Lambda$CDM between CMB and BAO datasets originates from their diverging inferences of the relative distances between the redshifts probed most sensitively by BAO ($z \sim 0.5-1$) and the redshift of  recombination ($z_*\sim 1100$), once the standard ruler of the acoustic scale has been calibrated by the CMB.  
This standard ruler is defined by the sound horizon 
\begin{equation}
    r(z_i) = \int_{z_i}^\infty \frac{dz}{H(z)} c_s(z) \,,
\end{equation}
where $c_s$ is the sound speed of the photon-baryon fluid.
For the CMB, the sound horizon is evaluated at recombination
$\rev{r_s}=r(\rev{z_s})$, and then its measured angular size $\rev{\theta_s}$ determines the comoving angular diameter distance
\begin{equation}
    D_M(z_i) = \int_0^{z_i}\frac{dz}{H(z)} \,,
\end{equation}
specifically $\rev{D_s \equiv D_M(z_s)}= \rev{r_s}/\rev{\theta_s}$. 
Here and throughout, we assume a spatially flat cosmology.
For BAO, the sound horizon is evaluated at the end of the drag epoch $r_d=r(z_d)$, and angular measurements of the transverse BAO scale determine $D_M(z)/r_d$\,, whereas redshift measurements of the radial BAO determine $D_H(z)/r_d$, where $D_H(z)=1/H(z)$\,.   

In \cref{fig:bao_data}, we show the DESI DR2 BAO measurements relative to the predictions of our fiducial  $\Lambda$CDM model.
The derived variable $D_V(z) \equiv 
[z D_M^2(z) D_H(z)]^{1/3}$ gives a visually useful combination of the BAO measurements $D_M$ and $D_H$, which tend to be anticorrelated at the same redshift. 
Note that for the lowest redshift point \rev{($z=0.295)$}, only $D_V$ is used in the likelihood whereas for other redshifts only $D_M$ and $D_H$ are used. 
The CMB-BAO tension can be seen in \cref{fig:bao_data} where $D_M/r_d$ and $D_V/r_d$ are measured to be systematically $\sim 1-2\%$ lower than predicted by fiducial \lcdm across the redshifts best measured by BAO (see also Fig. 6 in \cite{DESI:2025zgx}, where the best fit to BAO only is quantified). 
This tension can be reduced either by changing the calibrated $r_d$ or the relative distances $D_M(z)/\rev{D_s}$ or both.

\renewcommand{\arraystretch}{1.2} 
\begin{table*}[t]
    \centering
    \begin{tabular}{|l|c|c|c|c|c|c|c|c|c|}
        \hline
        \multicolumn{10}{|c|}{\textbf{CMB + BAO}} \\
        \hline
        \textbf{Model} & $\boldsymbol{\Delta\chi^2}$ & $\boldsymbol{\Delta\chi^2_{\rm CMB}}$ & $\boldsymbol{\Delta\chi^2_{\rm BAO}}$ & $\boldsymbol{r_d}$\,{\footnotesize [Mpc]} & $\boldsymbol{\fede}$ & $\boldsymbol{H_0}$ & $\boldsymbol{S_8}$ & $\boldsymbol{\Omega_m}$ &
        $\boldsymbol{D_M(0.8)/r_d}$\\
        \hline
        Fid. $\Lambda$CDM & \rev{16.9} & \rev{-5.9} & \rev{22.8} & \rev{147.00} & 0 & \rev{67.15} & \rev{0.84} & \rev{0.318} & \rev{19.59} \\
        $\Lambda$CDM & 0 & 0 & 0 & \rev{147.54} & 0 & \rev{68.16} & \rev{0.82} & \rev{0.303} & \rev{19.37} \\
        EDE & \rev{-9.4} & \rev{-7.4} & \rev{-1.9} & \rev{142.51} & \rev{0.09} & \rev{70.87} & \rev{0.83} & \rev{0.300} & \rev{19.32}\\
        \hline
    \end{tabular}
    \caption{
        Comparison of $\dchisq$ values 
         for CMB+BAO data
        and select parameters 
        of interest for $\Lambda$CDM tensions
        for various cosmological models. 
        Here \lcdm and EDE are optimized to CMB+BAO data, while fiducial \lcdm is simply evaluated at the ``Fid.~\lcdm'' cosmology in \cref{tab:pl18}. 
        We always report $\Delta\chi^2$ relative to best-fit $\Lambda$CDM for the dataset combination in question but include the difference with the fixed fiducial $\Lambda$CDM absolute numbers in \cref{tab:pl18} so that the fits can be assessed across different combinations.  
        Comparing the first and last rows shows that EDE is a better absolute fit to the CMB alone than fiducial $\Lambda$CDM, which is itself the best \lcdm fit to the CMB. 
    }
    \label{tab:cmb_bao}
\end{table*}

The calibration of $r_d$ is model-dependent. 
The presence of additional components like EDE that contribute to $H(z)$ before $z_d$ modify the resultant $r_d$. 
EDE specifically raises $H(z)$ and lowers $r_d$ and $\rev{r_s}$ when all else is fixed, so as to increase the CMB-inferred $H_0$ and reduce the Hubble tension. 
Since $\rev{\theta_s}=\rev{r_s}/\rev{D_s}$ is effectively fixed by the extremely precise CMB measurement, and lowering the inferred $\rev{D_s}$ by raising $H_0$ through the enhanced relative contribution of dark energy  has more of an impact at low vs.~high redshift, 
$D_M(z)/r_d$ is also lowered compared with $\Lambda$CDM (cf.~\cref{tab:cmb_bao} values for $\Omega_m=1-\Omega_\Lambda$).

\begin{figure}
    \centering
    \includegraphics[width=\linewidth]{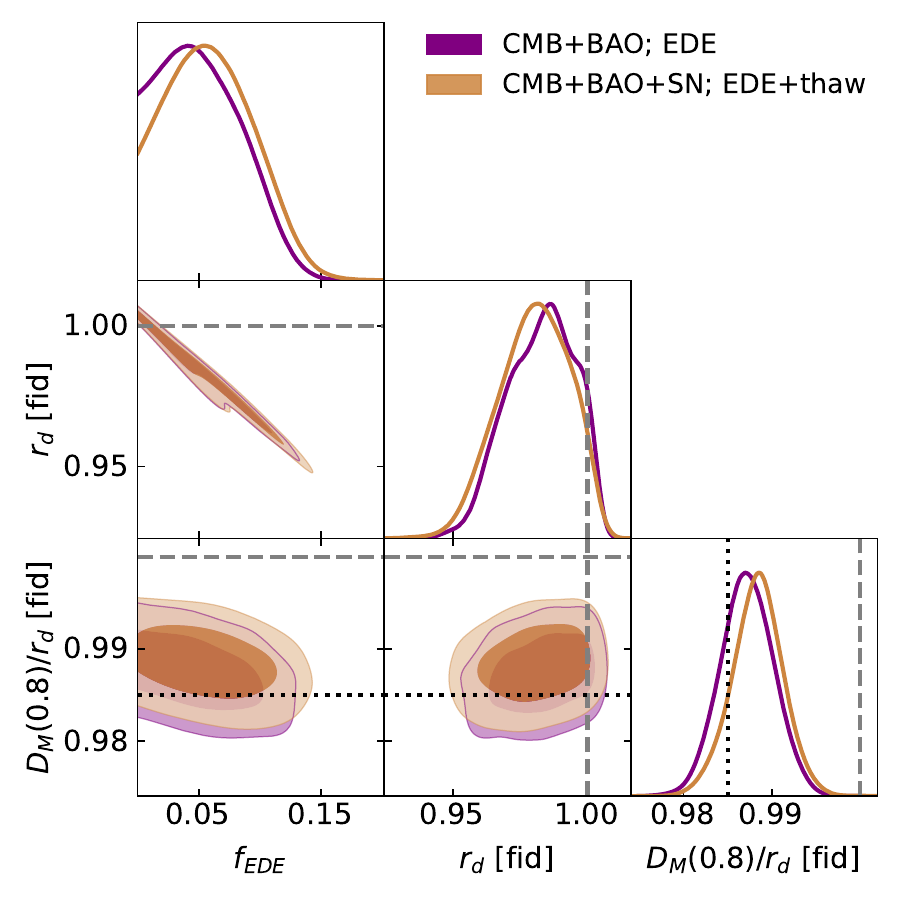}
    \caption{Posterior distributions for the CMB-BAO tension parameters 
    $D_M(z=0.8)/r_d$ and $r_d$ \rev{as the ratio to  those in the fiducial \lcdm cosmology, denoted by [fid]}, and the EDE parameter $\fede$, 
    for CMB+BAO; EDE and CMB+BAO+SN; EDE+thaw.  
    Gray dashed lines mark the fiducial $\Lambda$CDM values that best fit the CMB only. 
    The black dotted line shows the  BAO-only preference taken from Ref.~\cite{DESI:2025zgx} in their Fig. 6 (see also our \cref{fig:bao_data}). 
    Note that while the posterior including SN data also adds a thawing component besides EDE, the resolution of the CMB-BAO tension from EDE remains largely the same.}
    \label{fig:DM_rd_fid}
\end{figure}

In \cref{fig:DM_rd_fid}, we show that when raising the EDE fraction $\fede$\,, both $r_d$ and $D_M(z=0.8)/r_d$ decrease.   Here we have chosen $z=0.8$ as a rough proxy for the effective redshift of the CMB-BAO tension following Ref.~\cite{Liu:2025bss}.  Note that we choose $D_M(0.8)/r_d$ rather than the usual $H_0 r_d$ since the latter corresponds to BAO observables at $z=0$, which are not directly measured and depend more sensitively on the specific choice of late dark energy model (here a cosmological constant, see \cref{sec:sn} for exploration of other late dark energy parameterizations).  

\begin{figure}
    \centering
    \includegraphics[width=\linewidth]{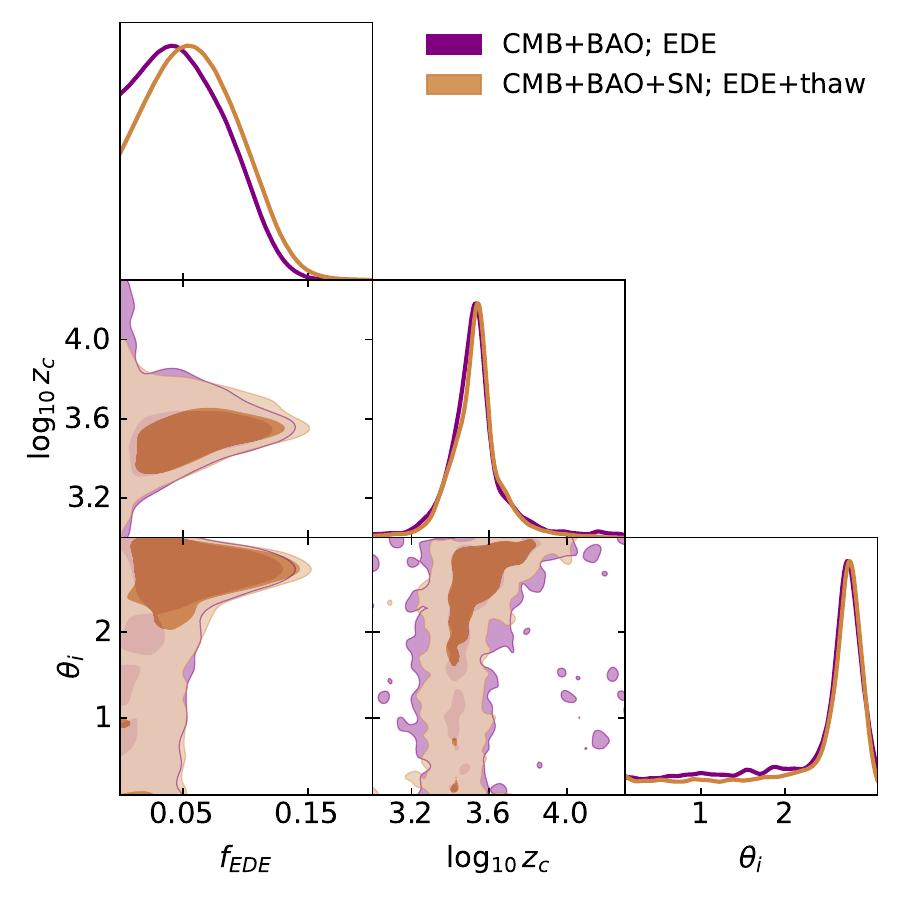}
    \caption{Posteriors of EDE parameters for CMB+BAO; EDE in purple and CMB+BAO+SN; EDE+thaw in orange.  The preferred EDE parameters remain largely unchanged regardless of the addition of low-redshift effects from SN and the thawing component. The widening of $z_c$ constraints at low $\fede$ leads to non-Gaussian posteriors in other parameters such as $r_d$ (see \cref{fig:DM_rd_fid}). }
    \label{fig:ede_posteriors}
\end{figure}

At lower values of \rev{$\fede\lesssim 0.08$} in \cref{fig:DM_rd_fid}, the posteriors have additional \rev{non-Gaussian} features especially visible in the correlation with $r_d$\,.  In \cref{fig:ede_posteriors}, we show that these correspond to the opening up of a wider range of $z_c$ and $\theta_i$, arising from the aforementioned prior-volume effects in EDE.  As we shall see next, these regions are 
distinct from the best-fit solution revealed by the profile likelihood.   

\begin{figure}
    \centering
    \includegraphics[width=\linewidth]{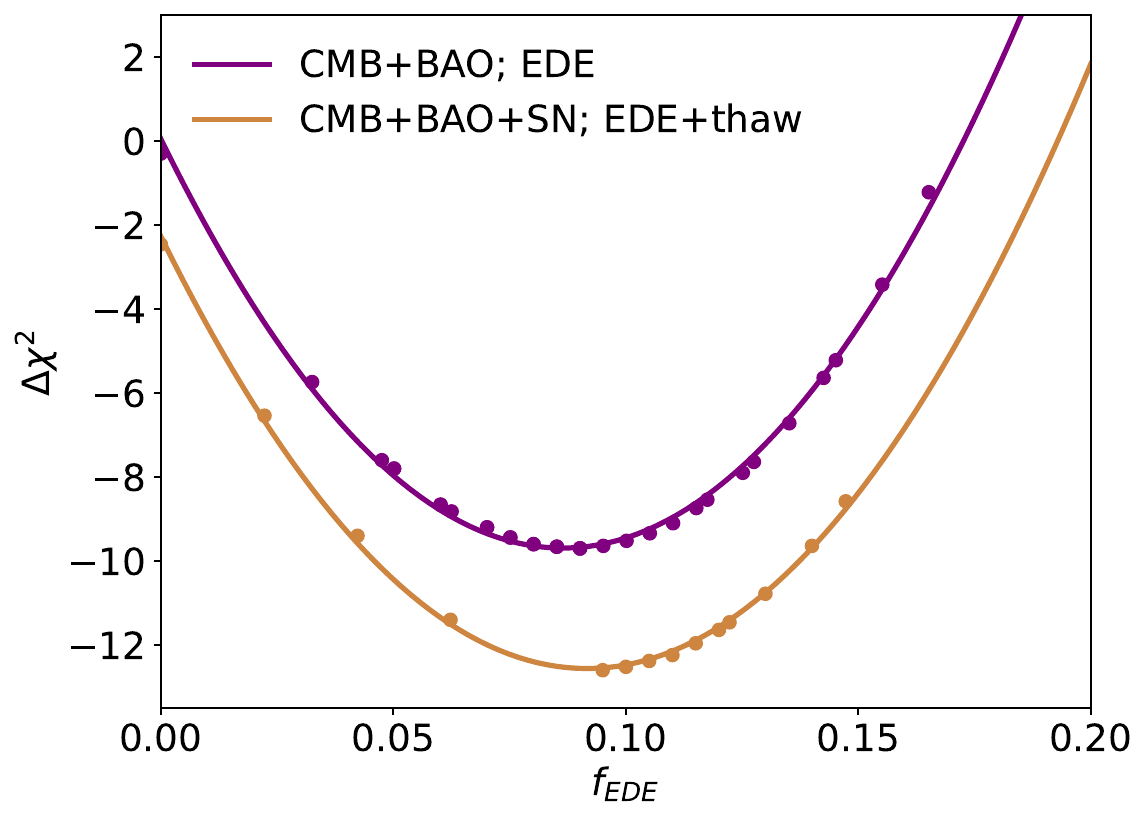}
    \caption{Profile likelihoods for the EDE fraction parameter $\fede$ for the CMB+BAO; EDE and CMB+BAO+SN; EDE + thaw data/model combinations. Values of $\Delta \chisq$ are relative to the best-fit \lcdm model for that data set.
    Both cases show substantial improvement over \lcdm at their  minima \rev{($\Delta\chi^2=-9.4$ and $-12.6$}, respectively), with the latter receiving an improvement (independent of $\fede$) in $\Delta\chi^2_{\rm SN}$ from the thawing quintessence model. The evaluation points are fitted with a quadratic curve and show that the prominent non-Gaussianity seen in the \rev{EDE parameter posteriors} in \cref{fig:ede_posteriors} are prior-volume artifacts.
    The offset between the cases reflects the raising of the distance modulus at $z<0.1$ through the thawing quintessence component.}
    \label{fig:fede_h0_omega_m_profile}
\end{figure}

In \cref{fig:fede_h0_omega_m_profile}, 
we show the profile likelihood for the $\fede$ parameter in the CMB+BAO; EDE data/model combination as the purple line. It is clear from comparing the profile likelihood to the one-dimensional posterior probability of $\fede\rev{=0.054^{+0.031}_{-0.047}}$
shown in \cref{fig:DM_rd_fid,fig:ede_posteriors} that: 1) the maximum marginalized posterior probability value for $\fede$ is being pulled down (compared to the profile value \rev{$\fede=\rev{0.09}\pm 0.03$ where $\pm$ refer to $\Delta\chi^2=\pm 1$ as estimated from a quadratic fit to the profile, nominally a $3\sigma$ preference}) by prior-volume effects, and 2) the profile likelihood is nearly quadratic, in contrast to the clear non-Gaussianity of the posterior.

The maximum-likelihood values of the parameters relevant for CMB+BAO tension  and the $\chi^2$ relative to $\Lambda$CDM---both in total and broken down by data set---for this data/model combination are given in \cref{tab:cmb_bao}.\footnote{\rev{Where they overlap, our results agree with Ref.~\cite{SPT-3G:2025vyw}\,v2, e.g. the largest differences here in $\Delta\chi^2$ between CMB+BAO EDE and $\Lambda$CDM  of order 1, due to their use of a $\tau$ prior instead of lowE data and Halofit instead of HMCode.}}
The best-fit EDE model improves the fit over \lcdm\ by $\Delta\chi^2 =-\rev{9.4}$ (with the addition of three free parameters, making the change in the Akaike Information Criterion $\Delta$AIC $=-\rev{3.4}$).

\Cref{tab:cmb_bao} also includes parameter and $\dchisq$ values for the fiducial $\Lambda$CDM model which is the best fit to CMB data alone.   Here we see that  the best-fit $\Lambda$CDM model for CMB+BAO must compromise on $\Delta\chi^2_{\rm CMB}$ to reduce the very large BAO penalty \rev{$\Delta\chi^2_{\rm BAO}=22.8$}.
The improvement of EDE over $\Lambda$CDM
is largely from the CMB with $\Delta\chi^2_{\rm CMB} =-\rev{7.4}$\,, of which \rev{$-5.7$} comes from primary anisotropy and the rest from lensing.
The EDE model is in fact a \rev{slightly} better fit to the CMB than even fiducial $\Lambda$CDM, where
$\Delta\chi^2_{\rm CMB}=\rev{-5.9}$ (with \rev{$-5.0$} coming from primary anisotropy),
despite fiducial $\Lambda$CDM, but not EDE, being optimized to fit CMB only.

\begin{figure*}[ht] 
    \centering
    \includegraphics[width=\textwidth]{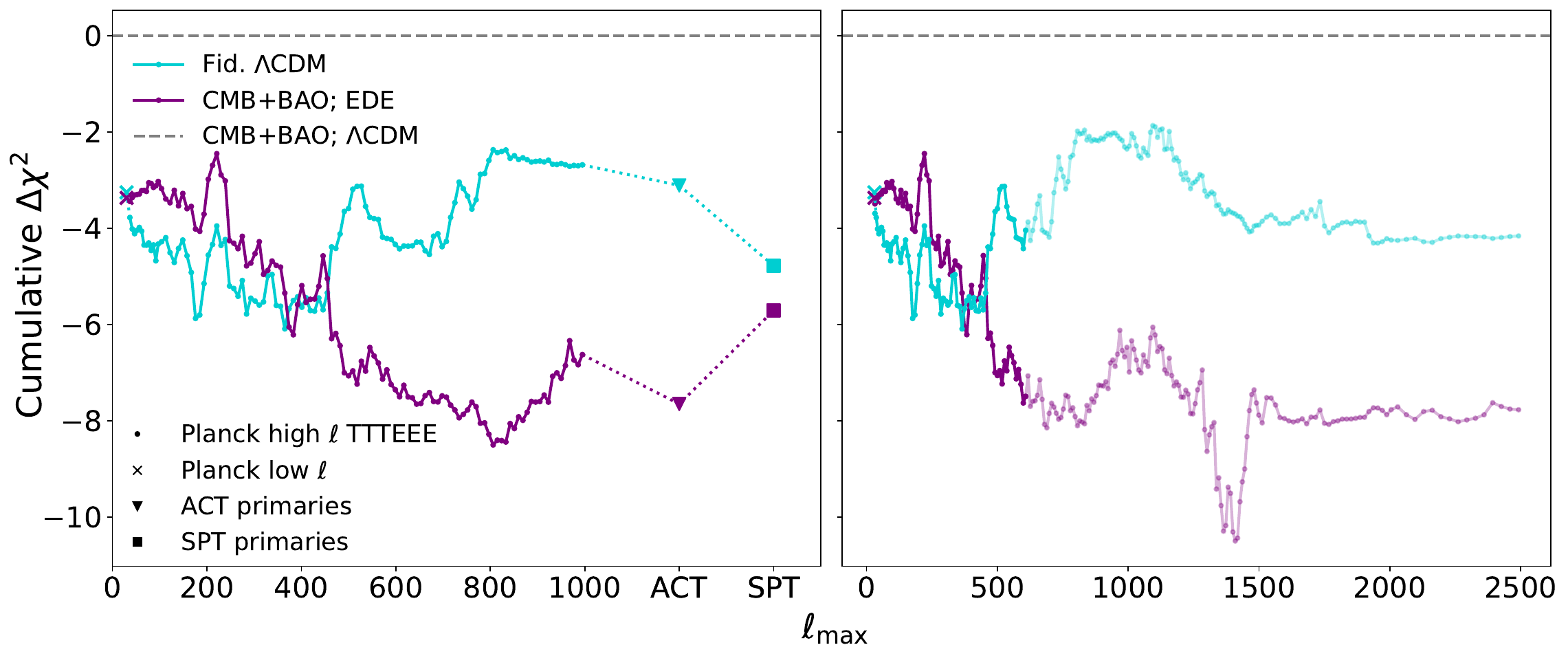}
    \caption{
    \rev{\textit{Left}:
    Cumulative contributions to $\Delta\chi^2$ for primary CMB anisotropy as a function of the maximum multipole $\ell_{\rm max}$ in the \planck, ACT and SPT datasets. The CMB+BAO tension-resolving EDE model and CMB-only optimized $\Lambda$CDM model are plotted relative to CMB+BAO $\Lambda$CDM.  Improvement at the lowest $\ell_{\rm max}$ in both reflect their lower value of $\tau$ and the EDE improvements between $200-600$ mainly reflect its lower $C_\ell^{TE}$ spectrum. Only \planck TT contributes for $600 < \ell_{\rm max} < 1000$.
    \textit{Right}: Cumulative $\Delta\chi^2$ for all of the \planck $\ell$-range for the same models as the left plot.  Despite not being readjusted for \planck TE,EE $\ell_{\rm max}>600$ or TT $\ell_{\rm max}>1000$, the preference for EDE grows to $\Delta\chi^2 \sim -7.8$ and is $-3.6$ better than fiducial $\Lambda$CDM.  These further contributions are not included in our fiducial $\Delta\chi^2$ analysis since they are correlated with ACT.}
    }
    \label{fig:cumulative}
\end{figure*}

\rev{
Fig.~\ref{fig:cumulative} ({left}) decomposes the improvement of EDE over $\Lambda$CDM  from primary anisotropy into its cumulative contributions to $\dchisq$ ordered in increasing $\ell_{\rm max}$.  We also compare the fiducial $\Lambda$CDM to this CMB+BAO optimized $\Lambda$CDM to provide a reference for how BAO data cause compromises in the CMB.
At the lowest $\ell_{\rm max}=30$, the improvements in both come from the lowE likelihood, with  {$(\Delta\chi^2_{\rm EDE}-\Delta\chi^2_{\rm \Lambda CDM})_{\rm lowE} ={-2.6}$ and {$(\Delta\chi^2_{\rm fid}-\Delta\chi^2_{\rm \Lambda CDM})_{\rm lowE} ={-4.0}$}}.
This comes mainly from the higher optical depth $\tau=0.0673$ in $\Lambda$CDM to reduce CMB+BAO tension over what the CMB  alone prefers in fiducial $\Lambda$CDM ($\tau=0.0555$, see \cite{{Sailer:2025lxj,Jhaveri:2025neg}}).  Since EDE alleviates CMB+BAO tension, \rev{its value for $\tau=0.0611$ approaches the fiducial value} (see Tab.~\ref{tab:pl18}). 
}

\rev{
In the intermediate range of $\ell \sim 400-600$, the EDE model performs better than both $\Lambda$CDM and fiducial $\Lambda$CDM mainly due to the TE spectrum and the ability of EDE to lower it coherently in this range \cite{Lin:2020jcb}.   This preference of EDE over fiducial $\Lambda$CDM from CMB primaries of $(\Delta\chi^2_{\rm EDE}-\Delta\chi^2_{\rm fid})_{\rm prim} \sim -4$ holds out through the \planck $\ell<1000$ range and the inclusion of ACT primary data but is reduced to near equivalence $(\Delta\chi^2_{\rm EDE}-\Delta\chi^2_{\rm fid})_{\rm prim} ={-0.9}$ once SPT primary data is also included.  Nonetheless the total $(\Delta\chi^2_{\rm EDE})_{\rm prim}= {-5.7}$ from CMB primaries is a more significant improvement over $\Lambda$CDM which must compromise the primary fit to reduce the tension.
}

\rev{Finally, Fig.~\ref{fig:cumulative} ({right}) shows the extension of the cumulative $\Delta\chi^2$ where ACT and SPT data are replaced by the full \planck dataset, i.e. including TE and EE for $\ell_{\rm max}>600$ and TT for $\ell_{\rm max}>1000$.  The feature around $\ell \sim 1400$ is due to 
EDE fitting the oscillatory residuals of $\Lambda$CDM that are related to the lensing anomaly in $\Lambda$CDM (see \cite{Lin:2020jcb}).
Note that the model parameters themselves are {\it not} reoptimized for this change and are identical to those in the left panel.  Despite this fact, the improvement of EDE over $\Lambda$CDM and fiducial $\Lambda$CDM increases to \rev{$\Delta\chi^2 \sim -7.8$ and $\sim -3.6$} respectively.} 

\rev{The lensing improvement of EDE over $\Lambda$CDM of \rev{$(\dchisq_{\rm EDE})_{\rm lens}=-1.7$ (not included in Fig.~\ref{fig:cumulative})} comes from two parameters. Compromising between CMB and BAO in \lcdm lowers $\omega_m$, which suppresses the predicted lensing amplitude. Under EDE, $\omega_m$ is restored back to its fiducial \lcdm level or higher. A higher $n_s$ further boosts lensing.}
This improvement is especially noticeable at high $\ell$ in the lensing potential power spectrum where SPT lensing measurements dominate. 

\begin{figure}
    \centering
    \includegraphics[width=\linewidth]{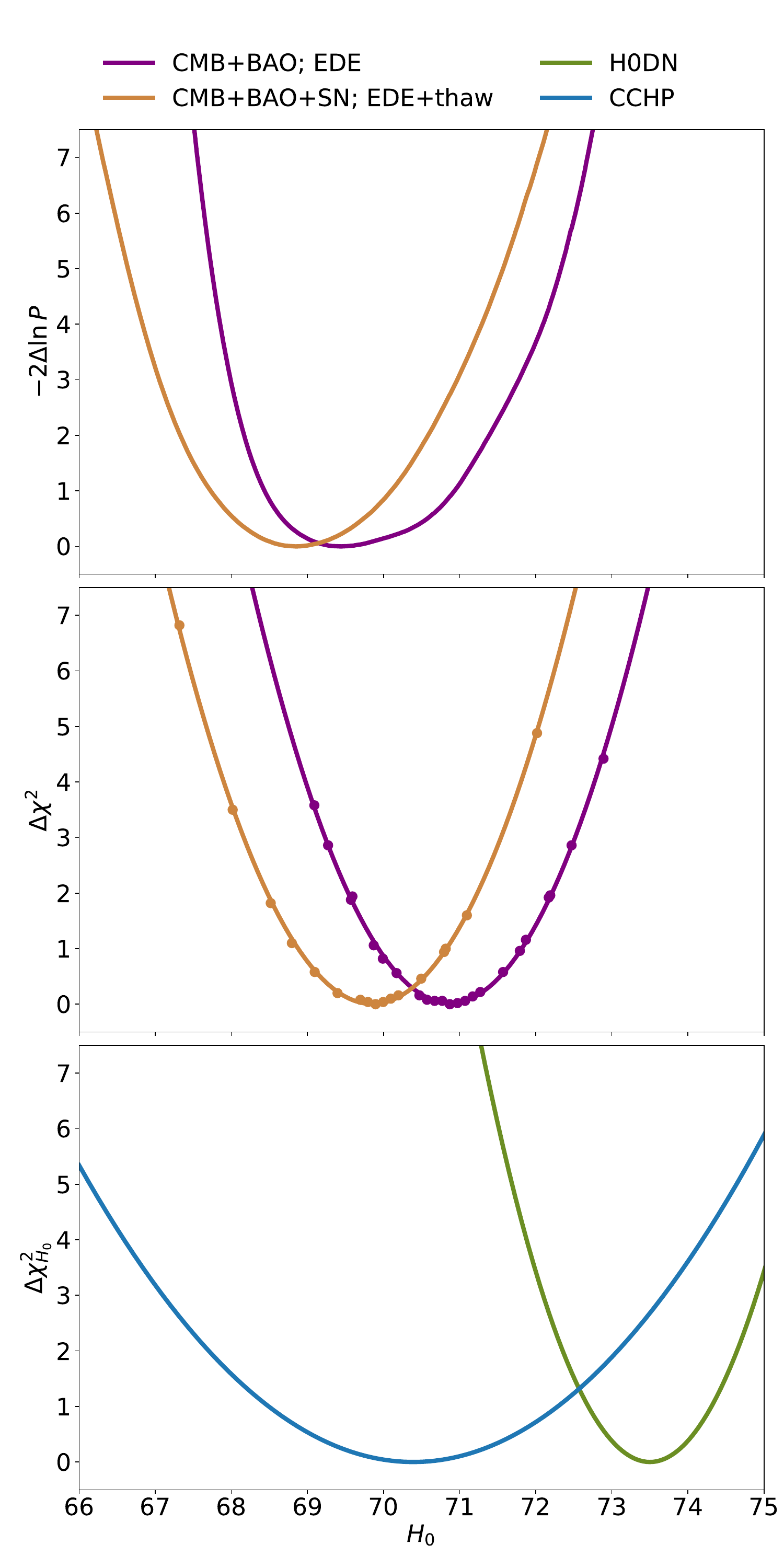}
    \caption{
    We show three different sets of 1D probability distributions for $H_0$, all recast as $\dchisq$ curves.
    \textit{Top}: Marginalized posterior probabilities $P$ for $H_0$ from the same dataset combinations and models as in Figs. \ref{fig:DM_rd_fid} through \ref{fig:fede_h0_omega_m_profile}, converted to $-2\Delta \ln P$ for comparison to $\Delta\chi^2$. 
    \textit{Middle}: Profile likelihoods in $H_0$, \rev{with evaluated points fitted by a quadratic,} for the same dataset combinations and models but normalized to zero at the  minimum for convenience of comparison. 
    \textit{Bottom}: the H0DN and CCHP constraints on $H_0$ converted to $\Delta\chi_{H_0}^2$
    using the quoted errors.
    }
    \label{fig:h0_profiles}
\end{figure}

Besides this mitigation of the CMB-BAO tension at no cost to $\dchisq_{\rm CMB}$\,, EDE also simultaneously reduces the Hubble tension. 
We explore this additional benefit in more detail in \cref{fig:h0_profiles}, with panels showing the marginalized posterior probability $-\Delta\ln P$ of $H_0$ interpreted as an effective $\Delta\chi^2$ profile relative to its minimum value (top panel),
the profile likelihood $\Delta\chi^2$ of $H_0$ (also relative to the minimum unlike other plots for ease of comparison), 
and the H0DN and CCHP $\Delta\chi^2_{H_0}$  as parabolas constructed from their reported errors (bottom panel). 

We first note that, comparing the top and middle panels of \cref{fig:h0_profiles}, the maximum-posterior-probability \rev{$H_0 = 69.83^{+0.89}_{-1.2}$ and maximum-likelihood values of $H_0=70.87\pm 0.94$ (where again $\pm$ refers to $\Delta\chi^2=\pm 1$)} are offset. This can be understood as the same prior-volume effect discussed above, 
combined with the $H_0$-$\fede$ correlation;
i.e., the expansion of prior volume at low values of $\fede$ biases the posteriors toward $\fede=0$, which in turn biases the $H_0$ posteriors low.

In terms of the Hubble tension, using the above profile error estimates combined with those of the local distance ladder measurements, we can evaluate the usual tension metric.
This leads to a difference-in-mean tension statistic of \rev{$2.1 \sigma$ and $0.2 \sigma$} to H0DN and CCHP, respectively. As advertised, the \rev{best-fit CMB+BAO EDE model  efficiently reduces} the Hubble tension 
from the nominal \rev{$7.1\sigma$ and $1.6\sigma$ in $\Lambda$CDM  \cite{SPT-3G:2025bzu}}, while also significantly reducing the CMB-BAO tension. 

On the other hand, these EDE solutions to the CMB-BAO (and $H_0$) tension potentially exacerbate other tensions. 
Most importantly for this work, by reducing $\Omega_m$, these EDE solutions drive $D_M$ at $z\lesssim 0.5$ even lower and potentially exacerbate the tension with SN, which favor the opposite trend for low-to-high-$z$ distance moduli.  We examine the extent to which SN block the EDE solution to the CMB-BAO(-$H_0$) tension in the next section.

\section{SN and Late Dark Energy}
\label{sec:sn}

In this section, we add SN data to the CMB+BAO data and investigate: 1) the degree to which the SN data disfavors the EDE solution to the CMB-BAO(-$H_0$) tension, and 2) the ability of a combination of EDE and a non-phantom late dark energy model to resolve the three-way CMB-BAO-SN tension, relative to $w_0 w_a$, and whether this combination preserves the Hubble-tension-alleviating properties of EDE alone.

\subsection{Fiducial Dovekie SN dataset}

\begin{figure}
    \centering
    \includegraphics[width=\linewidth]{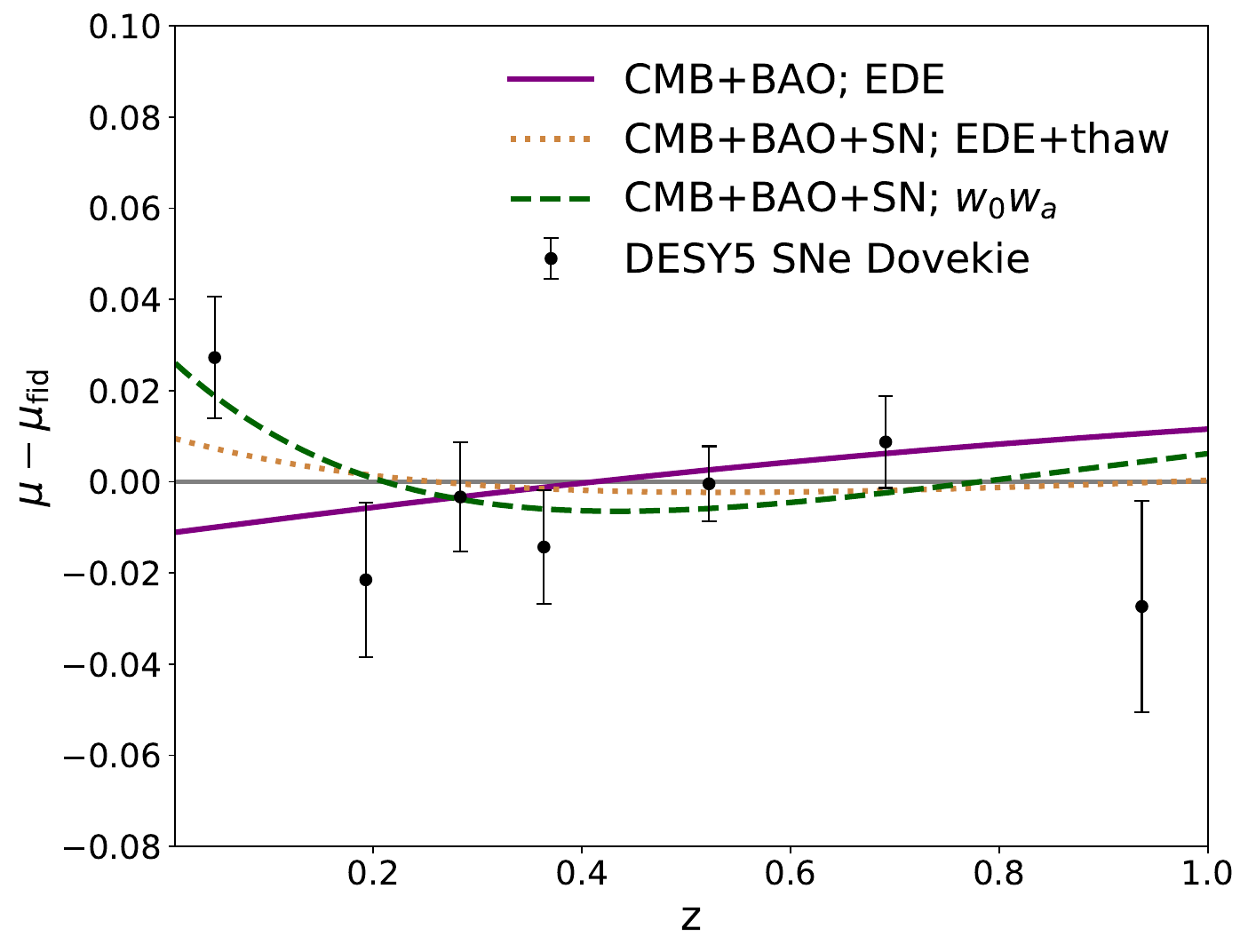}
    \caption{SN distance modulus data (relative to the values predicted in our fiducial \lcdm model) compared with various models.  
    The CMB+BAO; EDE (solid purple) case is not fit to SN and hence the $z\sim 0.1$ bin is not a good fit,  whereas 
    CMB+BAO+SN; EDE+thaw (dotted orange) fits a thawing quintessence component to accommodate this data.   
    The $w_0w_a$ model (dashed green) also fits the data but only through phantom crossing.  Note that the vertical offset of the curves is arbitrary due to the unconstrained SN absolute magnitude and is adjusted to best fit the points for visualization. }
    \label{fig:SN}
\end{figure}

We begin with our fiducial choice of SN datasets, DES \rev{Dovekie, which poses the most up-to-date SN test of the CMB-BAO EDE resolution}. 
Supernova magnitudes $m$ measure the distance modulus $\mu$ as 
\begin{equation}
    \mu = m-M =5 \log_{10}\left[\frac{ (1+z)D_M(z)}{{1 {\rm Mpc}}}\right]+25 \,,
\end{equation}
where the absolute magnitude $M$ is optimized as a nuisance parameter for each cosmological model in the absence of distance-ladder calibrations or $H_0$ measurements, and $m$ is corrected for peculiar velocities (see \cref{sec:methodology}).  As shown in \cref{fig:SN}, the EDE resolution of the CMB-BAO tension lowers the SN distance modulus at $z\lesssim 0.1$ vs.~higher redshifts, while the SN data prefer the opposite trend (which is the basis of the preference from SN for dynamical dark energy models when combined with other data sets).

\renewcommand{\arraystretch}{1.2} 
\begin{table*}[]
\centering
\vspace{0.5em}
\small
\setlength{\tabcolsep}{3pt}
\begin{tabular}{|l|c|c|c|c|c|c|c|c|c|c|c|}
    \hline
    \multicolumn{12}{|c|}{\textbf{CMB + BAO + SN}} \\
    \hline
    \textbf{Model} & $\boldsymbol{\Delta\chi^2}$ & $\boldsymbol{\Delta\chi^2_{\rm CMB}}$ & $\boldsymbol{\Delta\chi^2_{\rm BAO}}$ & $\boldsymbol{\Delta\chi^2_{\rm SN}}$ & $\boldsymbol{\fede}$ & $\boldsymbol{H_0}$ & $\boldsymbol{S_8}$ & $\boldsymbol{w_0}$ & $\boldsymbol{w_a}$ &
    $\boldsymbol{H_0r_d}$ &
    $\boldsymbol{D_M(0.8)/r_d}$ \\
    \hline
    Fid. $\Lambda$CDM & \rev{14.3} & \rev{-5.1} & \rev{21.7} & \rev{-2.3} & 0 & \rev{67.15} & \rev{0.84} & -1 & 0 & \rev{9871} & \rev{19.59} \\
    $\Lambda$CDM & 0 & 0 & 0 & 0 & 0 & \rev{68.07} & \rev{0.82} & -1 & 0 & \rev{10042} & \rev{19.39}\\
    EDE & \rev{-8.5} & \rev{-7.3} & \rev{-1.9} &  \rev{0.78} & \rev{0.08} & \rev{70.49} & \rev{0.83} & -1 & 0 & \rev{10077} & \rev{19.35} \\
    thaw & \rev{-2.5} & \rev{1.8} & \rev{-0.6} & \rev{-3.6} & - & \rev{67.27} & \rev{0.82} & \rev{-0.94} & \rev{-0.09} & \rev{9927} & \rev{19.40} \\
    EDE+thaw & \rev{-12.6} & \rev{-5.0} & \rev{-3.7} & \rev{-3.9} & \rev{ 0.10} & \rev{69.90} & \rev{0.84} & \rev{-0.93} & \rev{-0.11} & \rev{9945} & \rev{19.34}\\
    $w_0w_a$ & \rev{-15.8} & \rev{-6.5} & \rev{-4.4} & \rev{-4.8} & 0 & \rev{67.43} & \rev{0.83} & \rev{ -0.80} & \rev{-0.76} & \rev{9923} & \rev{19.24} \\
    \hline
\end{tabular}
    \caption{
    Comparison of best fits  and select parameters  across various cosmological models for combined fits to CMB+BAO+SN data, except the fiducial \lcdm cosmology (parameters shown in \cref{tab:pl18}) which is optimized to just CMB data. 
    }
    \label{tab:cmb_bao_sn}
\end{table*}

Correspondingly, once SN data are included with just EDE, the overall best fit degrades slightly. 
Analogous to \cref{tab:cmb_bao}, \cref{tab:cmb_bao_sn} contains maximum-likelihood values of relevant parameters and the $\chi^2$ relative to \lcdm, and we see from that table that the preference for EDE over \lcdm falls marginally from $\Delta\chi^2 = -\rev{9.4}$ to $\Delta\chi^2 =-\rev{8.5}$ relative to the best fit \lcdm, and the values $\fede=\rev{0.08}$ and $H_0=\rev{70.49}$ are \rev{slightly smaller but remain qualitatively unchanged}. 
Moreover, the EDE fit to the CMB \rev{becomes an even better}
fit than the fixed fiducial \lcdm model which is optimized to CMB data only. 

On the other hand, the best-fit $w_0 w_a$ model performs significantly better with 
$\Delta\chi^2=-\rev{15.8}$ ($\Delta$AIC $=-\rev{11.8}$), where the largest improvement over EDE is due to the SN component, while the CMB-BAO tension reduction is similar to EDE \cite{SPT-3G:2025vyw}, albeit without the added benefit of reducing the Hubble tension. 
In this sense, $w_0 w_a$ can be considered a candidate for resolving the three-way CMB-BAO-SN tension on its own, but at the cost of phantom dark energy behavior.

If we instead include a component of calibrated thawing quintessence, EDE+thaw for short, the best fit improves to $\Delta\chi^2 =-\rev{12.6}$ ($\Delta$AIC $= -\rev{4.6}$). The combination of EDE+thaw is thus able to resolve the three-way tension with a fit that is nearly as good as $w_0 w_a$, albeit with two extra parameters.
\Cref{fig:bao_data} compares this extended model to BAO data and shows that it preserves the CMB+BAO EDE solution. 
This is also quantified by comparing $\Delta\chi^2_{\rm BAO}$ vs.~the fixed fiducial \lcdm model in \cref{tab:cmb_bao,tab:cmb_bao_sn}.

Correspondingly in \cref{fig:DM_rd_fid}, the CMB-BAO tension in the $D_M(0.8)/r_d$ posterior is only marginally higher with the addition of SN data and calibrated thawing.  
\rev{
Note that although $H_0r_d$ provides a useful quantity with which to measure tension between CMB and BAO under \lcdm, where CMB prefers lower $H_0r_d$ than BAO, 
this metric can be misleading for dynamical dark energy models. 
BAO data best constrain the distance-redshift relation in the range $z\gtrsim 0.5$, which is extrapolated down to $z=0$ to obtain $H_0r_d$ in a given model. 
Dynamical dark energy scenarios that provide good fits to BAO distances can have very different $H_0r_d$ than what BAO data prefer under \lcdm. 
Relying solely on $H_0r_d$ values can then make a decent fit spuriously look like a bad one.  Notice that in Tab.\ \ref{tab:cmb_bao_sn} models with a good $\Delta\chi^2_{\rm BAO}$ share  similarly low values of $D_M(0.8)/r_d$ but do not necessarily share the same high values of $H_0 r_d$ as they would in $\Lambda$CDM.   In particular EDE+thaw shares the low value of $H_0 r_d$ with fiducial $\Lambda$CDM, which has a very bad $\Delta\chi^2_{\rm BAO}$, even though it has a better $\Delta\chi^2_{\rm BAO}$ than either EDE or $\Lambda$CDM.}

From \cref{fig:bao_data}, this can be attributed to the upswing in $D_M(z\rightarrow 0)/r_d= z/H_0 r_d$ due to fitting the thawing component to SN data at redshifts below the lowest BAO point \rev{which therefore does not substantially degrade the fit to BAO}.
Hence, \rev{though} $H_0 r_d$ presents a good metric to capture tension within \LCDM, or models that change high-$z$ physics only, \rev{it does} not do so when considering dynamical dark energy models. 

We conclude that EDE+thaw performs comparably to $w_0w_a$ in fitting the CMB-BAO-SN data. Though EDE+thaw requires two more free parameters than $w_0 w_a$, it avoids the phantom dark energy behavior of $w_0 w_a$ and retains some (but not all) of the added benefit of reducing the Hubble tension. 
Raising the $z\lesssim 0.1$ distance modulus through quintessence lowers $H_0$, here to a best-fit value of $H_0=\rev{69.90{\pm 0.92}}$ (where $\pm$ again refers to $\Delta\chi^2=\pm 1$ from the best fit). Yet this is still a significant reduction in the Hubble tension. 
Repeating the $H_0$ difference in mean significance 
from the previous section, but using the CMB+BAO+SN,
we find that $H_0$ tension is reduced to
\rev{$2.9\sigma$ and $0.2\sigma$ for H0DN and CCHP respectively, down from the nominal $7.1\sigma$ and $1.6\sigma$ for $\Lambda$CDM but up from $2.1\sigma$ and $0.2\sigma$ in CMB+BAO EDE. Note that this slight increase in H0DN tension compared with CMB+BAO EDE mainly reflects the thawing component and is present in thawing-only models as well.} 

\renewcommand{\arraystretch}{1.2} 
\begin{table*}[ht]
\centering
\vspace{0.5em}
    \begin{tabular}{|l|c|c|c|c|c|c|c|c|c|c|c|}
    \hline
    \multicolumn{12}{|c|}{\textbf{CMB + BAO + SN ($z>0.1$)}} \\
    \hline
    \textbf{Model} & $\boldsymbol{\Delta\chi^2}$ & $\boldsymbol{\Delta\chi^2_{\rm CMB}}$ & $\boldsymbol{\Delta\chi^2_{\rm BAO}}$ & $\boldsymbol{\Delta\chi^2_{\rm SN}}$ & $\boldsymbol{\fede}$ & $\boldsymbol{H_0}$ & $\boldsymbol{S_8}$ & $\boldsymbol{w_0}$ & $\boldsymbol{w_a}$ & $\boldsymbol{H_0r_d}$ & $\boldsymbol{D_M(0.8)/r_d}$ \\
    \hline
    Fid. \lcdm & \rev{17.2} & \rev{-6.1} & \rev{22.9} & \rev{0.4} & 0 & \rev{67.15} & \rev{0.84} & -1 & 0 & \rev{9871} & \rev{19.59}\\
    \lcdm & 0 & 0 & 0 & 0 & 0 & \rev{68.17} & \rev{0.82} & -1 & 0 & \rev{10058} & \rev{19.37}\\
    EDE & \rev{-9.4} & \rev{-7.5} & \rev{-2.0} & \rev{0.07} & \rev{0.09} & \rev{70.89} & \rev{0.83} & -1 & 0 & \rev{10101} & \rev{19.32} \\
    EDE+thaw & \rev{-9.4} & \rev{-7.8} & \rev{-1.6} & \rev{0.06} & \rev{0.09} & \rev{71.01}& \rev{0.83} &\rev{-1.01} & \rev{0.002} & \rev{10117} & \rev{19.32} \\
    $w_0w_a$ & \rev{-8.0} & \rev{-7.2} & \rev{-2.9} & \rev{2.1} & 0 & \rev{67.83} & \rev{0.83} & \rev{-0.85} & \rev{-0.62} & \rev{9986} & \rev{19.25} \\
    \hline
    \noalign{\vskip 3mm}
    \hline
    \multicolumn{12}{|c|}{\textbf{CMB + BAO + SN (\rev{DESY5})}} \\
    \hline
    \textbf{Model} & $\boldsymbol{\Delta\chi^2}$ & $\boldsymbol{\Delta\chi^2_{\rm CMB}}$ & $\boldsymbol{\Delta\chi^2_{\rm BAO}}$ & $\boldsymbol{\Delta\chi^2_{\rm SN}}$ & $\boldsymbol{\fede}$ & $\boldsymbol{H_0}$ & $\boldsymbol{S_8}$ & $\boldsymbol{w_0}$ & $\boldsymbol{w_a}$ & $\boldsymbol{H_0r_d}$ & $\boldsymbol{D_M(0.8)/r_d}$\\
    \hline
    Fid. \lcdm & \rev{12.7} & \rev{-4.2} & \rev{20.7} & \rev{-3.8} & 0 & \rev{67.15} & \rev{0.84} & -1 & 0 & \rev{9871} & \rev{19.59} \\
    \lcdm & 0 & 0 & 0 & 0 & 0 & \rev{68.01} & \rev{0.82} & -1 & 0 & \rev{10028} & \rev{19.40}\\
    EDE & \rev{-8.0} & \rev{-7.0} & \rev{-2.1} & \rev{1.1} & \rev{0.08} & \rev{70.35} & \rev{0.83} & -1 & 0 & \rev{10062} & \rev{19.37}\\
    EDE+thaw & \rev{-17.3} & \rev{-3.3} & \rev{-4.6} & \rev{-9.4} & \rev{0.10} & \rev{69.32} & \rev{0.84} & \rev{ -0.89} & \rev{-0.17} & \rev{9854} & \rev{19.36} \\
    $w_0 w_a$ & -\rev{22.7} & \rev{-5.7} & \rev{-6.5} & \rev{-10.5} & 0 & \rev{66.82} & \rev{0.84} & \rev{-0.75} & \rev{-0.88} & \rev{9834} & \rev{19.25} \\
    \hline
    \noalign{\vskip 3mm}
    \hline
    \multicolumn{12}{|c|}{\textbf{CMB + BAO + SN (Pantheon+)}} \\
    \hline
    \textbf{Model} & $\boldsymbol{\rev{\Delta}\chi^2}$ & $\boldsymbol{\Delta\chi^2_{\rm CMB}}$ & $\boldsymbol{\Delta\chi^2_{\rm BAO}}$ & $\boldsymbol{\Delta\chi^2_{\rm SN}}$ & $\boldsymbol{\fede}$ & $\boldsymbol{H_0}$ & $\boldsymbol{S_8}$ & $\boldsymbol{w_0}$ & $\boldsymbol{w_a}$ & $\boldsymbol{H_0r_d}$ & $\boldsymbol{D_M(0.8)/r_d}$\\
    \hline
    Fid. \lcdm & \rev{14.8} & \rev{-5.3} & \rev{22.0} & \rev{-1.9} & 0 & \rev{67.15} & \rev{0.84} & -1 & 0 & \rev{9871} & \rev{19.59} \\
    \lcdm & 0 & 0 & 0 & 0 & 0 & \rev{68.09} & \rev{0.82} & -1 & 0 & \rev{10045} & \rev{19.38} \\
    EDE & \rev{-8.7} & \rev{-7.4} & \rev{-2.0} & \rev{0.7} & \rev{0.08} & \rev{70.50} & \rev{0.83} & -1 & 0 & \rev{10082} & \rev{19.34} \\
    EDE+thaw & \rev{-11.7} & \rev{-5.5} & \rev{-3.2} & \rev{-3.0} & \rev{0.09} & \rev{69.93} & \rev{0.84} & \rev{-0.94} & \rev{-0.095} & \rev{9952} & \rev{19.35} \\
    $w_0 w_a$ & \rev{-13.1} & \rev{-6.3} & \rev{-4.2} & \rev{-2.5} & 0 & \rev{67.56} & \rev{0.83} & \rev{-0.83} & \rev{-0.66} & \rev{9945} & \rev{19.26} \\
    \hline
    \end{tabular}
    \caption{
    Comparison of best fits and select parameters across various cosmological models for combined fits to CMB+BAO+SN data incorporating alternative SN datasets (see text for discussion). 
    \textit{Top:} We remove all non-DES SNe from the \rev{Dovekie} SN sample, effectively removing SNe with $z<0.1$. 
    \textit{Middle:} We replace  \rev{the Dovekie with the DESY5} SN sample. 
    \textit{Bottom:} We replace  \rev{the Dovekie with the Pantheon+ uncalibrated SN sample.}
    }
\label{tab: sn_exploration}
\end{table*}

\subsection{Alternative SN datasets}

We have seen that the preference for late dark energy in our choice of CMB+BAO+SN data comes from the relatively high distance modulus at the lowest redshifts compared with the higher redshifts of the \rev{Dovekie} SN catalog. In this section, we explore the dependence of this preference on the specific choice of SN dataset, including how this choice impacts the overall data preference for EDE+thaw vs EDE and~$w_0 w_a$.  The alternative SN data sets are defined in \cref{sec:methodology}, and the results of these explorations are summarized in \cref{tab: sn_exploration}.

Our first test is to drop from the analysis any SNe in the \rev{Dovekie} data set that do not come from the DES survey itself, as these may be subject to different systematics \cite{Efstathiou:2024xcq}.
This also effectively cuts all SNe with $z<0.1$, so we call this alternate data SN ($z > 0.1$).
In this case, the block on the EDE-only solution to the CMB-BAO tension essentially disappears with the preference over \lcdm returning to $\Delta\chi^2\sim -\rev{9.4}$ (cf.~\cref{tab:cmb_bao}) with or without the additional thawing component, \rev{showing that the addition is not necessary}.
In fact both are then a better fit than $w_0 w_a$ where $\Delta \chi^2 = \rev{-8.0}$.  This shows that aside from the $z<0.1$ SN data, the EDE solution works remarkably well for all other aspects of the data and also has the highest $H_0 \approx  \rev{71}$. 

\rev{Next we consider the \rev{original DESY5} catalog.
The $w_0 w_a$ preference  was reported in Ref.~\cite{DES:2025sig} 
to change from  $3.2\sigma$ (Dovekie) to $4.2\sigma$ (DESY5) in their combination with CMB and BAO.} 
Correspondingly, in our combination, the preference for $w_0w_a$ over \lcdm \rev{increases} from $\Delta\chi^2=-\rev{15.8}$ to $-\rev{22.7}$. 
Furthermore, the EDE+thaw model is now  \rev{somewhat further from} the $w_0 w_a$ solution than in the baseline \rev{Dovekie} case, with $\chi^2_\mathrm{EDE+thaw} - \chi^2_{w_0 w_a} = \rev{5.4} $ instead of $\rev{3.2}$.
The best-fit thawing $w_0$ for EDE+thaw also \rev{rises from $-0.93$ to $-0.89$, further from a simple cosmological constant.}  In fact the improvement over EDE-only also \rev{rises to $-9.3$} for the one additional parameter $w_0$.

Finally the Pantheon+ SN data set also shows a smaller preference for $w_0w_a$ than DESY5 when combined with CMB+BAO \cite{DESI:2025zgx}.  
In this case, the difference in $\Delta\chi^2$ between $w_0w_a$ and EDE+thaw narrows to \rev{-1.4}.
The thawing component contributes \rev{$-3$} for the one extra parameter. 

In all of these SN variants, the best-fit $\fede\sim \rev{0.08-0.1}$ and $H_0$ varies from \rev{$\sim 69.32-71.01$}\, 
with the lower end representing cases where a thawing component is added to EDE, which \rev{improves the fit by varying degrees} at the cost of one extra parameter.
We conclude that both the need to add thawing quintessence to EDE and also the difference between the fits with $w_0w_a$ and EDE+thaw depend on the SN dataset and in particular the relative distances between SNe at $z<0.1$ and higher redshifts.   Conversely, the CMB-BAO(-$H_0$) resolution remains largely insensitive to the SN dataset and most of the preference of different datasets for dynamical dark energy can be absorbed into a non-phantom thawing component in all cases.  

\section{Discussion}
\label{sec:discussion}

We have extended the findings of other recent works that early dark energy provides a possible solution to tension between CMB and BAO distance measures while also reducing the Hubble tension, by using the most up-to-date CMB, BAO and local $H_0$ data and analyzing them in a manner that highlights  its robustness to late dark energy changes to distance measures.
Moreover, while EDE does not directly address deviations in the SN distance modulus at these lowest redshifts, we have demonstrated that any such preference can be absorbed into a component of thawing quintessence dark energy without the need for exotic phantom behavior. The solution to the CMB-BAO tension is largely immune to changes in the SN data set, unlike the $w_0w_a$ phantom solution.  

Specifically, we have shown that EDE relaxes the CMB-BAO tension both by changing the drag scale $r_d$---which also raises $H_0$ to a best-fit value of \rev{$70.49$}---and by lowering the BAO distance measure $D_M/r_d$ by the necessary $\sim 1-2\%$ amount to best fit the DESI DR2 measurements at $z\sim 0.8$. The overall improvement in CMB+BAO relative to \lcdm is $\dchisq =-\rev{9.4}$.

While this improvement is reduced to $\Delta\chi^2=-\rev{8.5}$ by including DES Dovekie SN (with the critical $z<0.1$ SNe), the fit is more than restored to $\Delta\chi^2=-\rev{12.6}$ by adding a normal thawing-quintessence component to accommodate the preference for evolving late dark energy.  While $w_0w_a$ still provides a larger improvement of $\dchisq=-\rev{15.8}$, including significant improvement in CMB-BAO tension (though not in the Hubble tension), EDE+thaw achieves its improvement without an exotic and potentially unstable component of dark energy.   Moreover with other SN datasets, \rev{this small difference varies}.
For example with Pantheon+, the $w_0w_a$ and EDE+thaw \rev{difference narrows and they fit comparably well}.

Finally, we have demonstrated that the EDE solution, with or without SN data and late dark energy, substantially reduces the Hubble tension. Comparing profile likelihoods of $H_0$ (which, unlike marginalized posteriors, do not bias $H_0$ low in EDE models) with probability distributions of local measurements, we find 
that with the Dovekie data and including a thawing quintessence component, our value of $H_0$ is within an equivalent \rev{$2.9 \sigma$ distance} of the H0DN measurement, compared with \rev{$7.1\sigma$} in $\Lambda$CDM,
and is completely compatible with the CCHP measurement.  Dropping SN data sets entirely, the tension with H0DN with EDE is an even smaller \rev{$2.1\sigma$} and alternate SN datasets would have variations that \rev{mainly reflect their varying preference for quintessence.} 

Data available in the near future will sharpen these comparisons and demonstrate decisively whether there is strong evidence for either early dark energy or evolving late dark energy, and whether a non-phantom solution is viable. In particular, new SN data from the Vera Rubin Observatory's Legacy Survey of Space and Time \cite{LSST:2008ijt} and the Nancy Grace Roman Space Telescope \cite{Dore:2019pld} will greatly increase the SN constraining power. At the same time, efforts to improve and standardize the calibration of the crucial $z < 0.1$ SNe are ongoing. In the area of CMB data, we can look forward to new data from the Simons Observatory Large-aperture Telescope \cite{SimonsObservatory:2018koc} and the release of the SPT-3G Ext-10k survey of one quarter of the sky \cite{SPT-3G:2024qkd}, which will improve CMB constraints on extensions to \lcdm by factors of two or more. Finally, techniques such as those used in \cite{SPT-3G:2024atg}, which reconstruct the unlensed CMB at the surface of last scattering, have the potential to resolve EDE-related features in the CMB power spectrum (particularly in polarization, see \cite{Lin:2020jcb}) that are washed out by lensing. For now, we conclude that a combination of EDE and non-phantom late dark energy is a viable candidate to resolve tensions among the CMB, BAO, and SN data while simultaneously improving the Hubble tension.

\acknowledgments 
We thank Graeme Addison and Lloyd Knox for useful conversations and Rayne Liu for help in SN binning and likelihood crosschecks.
T.J. and W.H. are supported by U.S.\ Dept.\ of Energy contract DE-SC0009924 and the Simons Foundation; T.J. and T.C. are supported by the National Science Foundation through awards OPP-1852617 and OPP-233248 under the South Pole Telescope Program.
T.K. is supported by the Kavli Institute for Cosmological Physics at the University of Chicago through an endowment from the Kavli Foundation. 
Computing resources were provided by the  University of Chicago  Research
Computing Center through the Kavli Institute for Cosmological Physics.
\rev{This project has received funding from the European Research Council (ERC) under the European Union’s Horizon 2020 research and innovation programme (grant agreement No 101001897).}

\vfill

\bibliographystyle{apsrev4-2}
\bibliography{references}

\end{document}